\documentclass[prl,aps,twocolumn,floatfix,showpacs,tightenlines,superscriptaddress,amsmath,amssymb]{revtex4}
\usepackage{amssymb,amsbsy,epsfig,color,graphicx,times}

\newcommand{\ket}[1]{|#1\rangle}
\newcommand{\bra}[1]{\langle#1|}

\newcommand{\eq}[1]{Eq.~(\ref{#1})}
\newcommand{\fig}[1]{Fig.~\ref{#1}}

\begin{document}



\title{Probing Quantum Frustrated Systems via Factorization of the Ground State}

\author{Salvatore M. Giampaolo}
\affiliation{Dipartimento di Matematica e Informatica, Universit\`a
degli Studi di Salerno, CNR-SPIN, CNISM, Unit\`a di Salerno, and INFN,
Sezione di Napoli - Gruppo Collegato di Salerno, Via Ponte don Melillo,
I-84084 Fisciano (SA), Italy}

\author{Gerardo Adesso}
\affiliation{School of Mathematical Sciences, University of Nottingham,
University Park,  Nottingham NG7 2RD, UK}

\author{Fabrizio Illuminati}
\affiliation{Dipartimento di Matematica e Informatica, Universit\`a
degli Studi di Salerno, CNR-SPIN, CNISM, Unit\`a di Salerno, and INFN,
Sezione di Napoli - Gruppo Collegato di Salerno, Via Ponte don Melillo,
I-84084 Fisciano (SA), Italy}
\affiliation{Corresponding author: illuminati@sa.infn.it}


\begin{abstract}
The existence of definite orders with simple periodicity in frustrated
quantum systems is related rigorously to the occurrence of fully factorized
ground states below a threshold value of the frustration. Ground state
separability thus provides a natural measure of frustration: Strongly frustrated
systems are those that cannot accommodate for unentangled and
mean field solutions. The exact form of the factorized ground states
and the critical frustration are determined for various classes of
non-exactly solvable spin models with different spatial ranges of the interactions.
For weakly frustrated systems, the existence of disentangling transitions
determines the range of applicability of mean field descriptions in
biological and physical problems such as stochastic gene expression
and the stability of long-period modulated structures.
\end{abstract}

\date{May 19, 2010}

\maketitle

\noindent {\bf Introduction.--} Interest in frustrated
quantum systems is due to the fact that they exhibit a large ground
state degeneracy, or quasi-degeneracy, associated to
complex structures of the quantum phase diagrams \cite{Diep}. Moreover,
frustrated spin models arise naturally in a variety of physical
situations, e.g., in the study of high-$T_c$ superconductivity \cite{Anderson,Mambrini},
long-period modulated structures of condensed matter \cite{ANNNI},
and biological systems with stochastic components \cite{Sasai}.
Unfortunately, rigorous results are available so far only
for ultra-simplified models \cite{Diep}, while numerical simulations
are challenging, because quantum Monte Carlo methods are not practical
for frustrated spin and fermion models \cite{Troyer}, and the density
matrix renormalization group is difficult to apply to systems with dimensionality
larger than one and/or periodic boundary conditions \cite{Schollwoek}.
Furthermore, criteria to quantify frustration and the classification
of "weakly" and "strongly" frustrated systems have been so far phenomenological
and/or numerical, with little or no rigorous physical and mathematical
insight \cite{Lhuillier}.

Recently, a method based on quantum information techniques \cite{GiampaoloIlluminati},
has allowed us to establish rigorously that
large classes of (generally non exactly solvable) frustration-free
quantum spin models admit totally disentangled ground states (GS) at
finite values of the interaction strengths and of the external
fields \cite{PrlLungo}. These factorized GS coincide
exactly with the mean field solutions and identify well defined
magnetic orders endowed with simple spatial periodicities.
In this Letter we build on that formalism 
to investigate the physics of quantum frustrated systems by showing that
GS separability is a measure of quantum frustration that discriminates quantitatively
regimes of weak and strong frustration. Indeed, we prove the existence of a
critical frustration (frustration threshold) below which fully factorized GS
are allowed and correspond to a definite magnetic orders with simple periodicity
for quantum spin models belonging to different universality classes and with
different types and ranges of interactions.
We then prove rigorously the existence of disentangling transitions in the GS in
the regime of weak frustration and identify the different magnetic orders and their
quantum phase boundaries. That, indeed, entanglement and separability can be used to
qualify and quantify frustration (and viceversa) can be intuitively understood by
observing that the presence of frustration tends to enhance correlations among the
constituents and thus to depress the possibility for the occurrence of separable (uncorrelated)
states. 
We also discuss some predictive consequences of these results on the study of complex
systems in physics and biology. Although in what follows, for ease of presentation,
the analysis is carried out in detail for one-dimensional systems, it is straightforward
to extend it and apply it in general to systems of arbitrary dimensionality.




Consider, without loss of generality, spin-$1/2$ systems for which frustration arises from
the simultaneous presence of competing antiferromagnetic exchange interactions of different
spatial range. The anisotropy $J_\alpha \ge 0$ ($\alpha=x,y,z$) in
the spin-spin coupling at sites $i$ and $j$ of the lattice with
distance $r=|i-j|$ is taken independent of $r$, and all couplings
are rescaled by a common, distance-dependent factor $f_r > 0$. The
general model Hamiltonian then reads
\begin{equation}\label{HamRange}
    H= \! \! \! \! \! \! \! \sum_{i,r\le r_{\max}} \! \! \! \! \! \! \! f_r(J_x S_i^x S_{i+r}^x\! \!+\!
    J_y S_i^y S_{i+r}^y\!\!+\!J_z S_i^z S_{i+r}^z\!)
     - h \! \sum_{i} S_i^z\! ,
\end{equation}
where $S_i^\alpha$ are the spin-$1/2$ operators at site $i$; $h$ is
the external magnetic field; and $r_{\max} > 1$ is the interaction
range, i.e.~the maximum distance between two spins with nonvanishing
coupling. Without loss of generality we assume $J_x \ge J_y$. This
general $XYZ$ Hamiltonian with interactions of arbitrary range
includes as sub-cases very many different models spanning several
classes of universality, such as the short- and long-range Ising,
Heisenberg, $XY$, $XX$, and $XXZ$ models.
We first recall briefly the basic findings on GS
factorization in frustration-free spin models \cite{PrlLungo}. The quantity
controlling GS factorization is the entanglement excitation energy
(EXE) $\Delta E$ \cite{GiampaoloIlluminati}. At every site
$k$ it is defined as
$\Delta E = \min_{\{U_{k}\}}\bra{G}U_k H U_k \ket{G}-\bra{G}H \ket{G}$.
Here $\ket{G}$ is the GS of the system and $U_k$ is any local
rotation acting on the spin at site $k$, i.e. a {\it single-spin
unitary operation} \cite{GiampaoloIlluminati}: $U_{k} \equiv
\bigotimes_{i \neq k} \textbf{1}_i \otimes 2 {O}_k $, where
$\textbf{1}_i$ is the identity operator on all the spins but the one
at site $k$, and ${O}_k$ is a generic Hermitian, unitary, and
traceless operator \cite{GiampaoloIlluminati}. For any
translationally invariant and frustration-free Hamiltonian $H$ such
that $[H,U_k] \neq 0$ $\forall U_k$, the vanishing of the EXE is a
necessary and sufficient condition for GS factorization
\cite{GiampaoloIlluminati}.
In fact, the minimization defining the EXE identifies an extremal
operation $\bar{U}_k$ at each site of the lattice and, therefore, a
global operator $\bar{U} \equiv \bigotimes_k \bar{U}_k$ that admits as
its own eigenstate the fully factorized (separable) state
$\ket{G_F}$:
\begin{equation}\label{Factorized}
\ket{G_F}=\prod_k \left[
\cos({\theta}_k/2)\ket{\uparrow_k}+e^{i{\varphi}_k}
\sin({\theta}_k/2)\ket{\downarrow_k}\right] \, ,
\end{equation}
where ${\theta}_k$ and ${\varphi}_k$ define the
direction of $\bar{U}_k$ in spin space. $\ket{G_F}$ is the exact GS if and only
if the EXE vanishes. 
Proceeding to investigate frustrated spin models by this method,
we consider first the simplest short-range interactions,
i.e. antiferromagnetic interactions that extend
up to nearest-neighbor (nn) and next-nearest-neighbor (nnn)
spins ($r_{\max}=2$).


\smallskip

\begin{figure}[t]
\includegraphics[width=8cm]{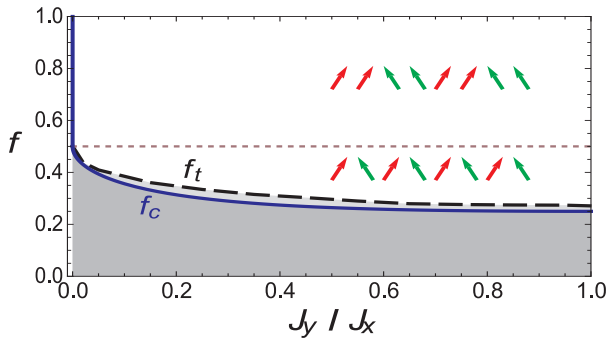}
\caption{(Color online) Analytical lower bound $f_c$ (solid blue
line) and exact numerical value of the frustration compatibility
threshold $f_t$ (dashed black line) as functions of the ratio
$J_y/J_x$. GS factorization occurs if and only if $f <
f_t$. For $f$ below the horizontal dotted line $f = 1/2$
the magnetic order is single-step
antiferromagnetic (SA), while for $f$ above it, it is dimerized
antiferromagnetic (DA). Therefore, no factorized GS
supports DA order, except at $J_y = 0$.}
\label{facground}
\end{figure}

\noindent {\bf Short-range models of frustrated antiferromagnets.--}
For $r_{\max}=2$, with ordering $f_1 > f_2 > 0$, in the Hamiltonian \eq{HamRange}
we can set, without loss of generality, $f_1 = 1$, so that the
parameter $f = f_2 / f_1 \in [0,1]$ quantifies the {\em degree of
frustration}: For $f=0$ the system is frustration-free, while for
$f=1$ the model is fully frustrated.
To verify the existence of a factorized GS we impose
minimization of the energy and the vanishing of the EXE
to determine the explicit expressions of ${\theta}_k$
and ${\varphi}_k$.
One finds that as long as $f < 1/2$ a factorized GS
exists and is associated to the single-step antiferromagnetic (SA)
order along the $x$ axis. This behavior is mirrored in the fact
that ${\varphi}_k = k \pi$, $\forall k$. Viceversa, as soon as $f
\ge 1/2$, the candidate factorized GS is associated to a dimerized
antiferromagnetic order (DA), corresponding to alternating local phases: $\varphi_{2k}= k \pi, \;
\varphi_{2k+1}=\varphi_{2k}$ (See Fig.~\ref{facground}).
The angle $\theta_k$ is site-independent: $\theta_k \equiv \theta$ $\forall k$,
and is the solution of
\begin{equation}
\label{e.theta}
\cos \theta =  \frac{2 h_F}{\mathcal{J}_z-\mathcal{J}_x} \; ,
\end{equation}
where $h_F$ stands for the factorizing field, i.e. the value (at
this stage, yet to be determined) of the external field at which the
GS becomes fully separable. The quantities $\mathcal{J}_\alpha$ are
the net interactions that express the coupling of the entire system
to a given spin, due to the presence of the external field. The net
interaction along the $z$-axis is independent of the magnetic order:
$\mathcal{J}_z = 2 J_z(1+f)$, while for the ones along $x$ and $y$
one has $\mathcal{J}_{x,y}=-2 (1-f) J_{x,y}$ in the presence of SA
order $(f < 1/2)$, and $\mathcal{J}_{x,y}=-2 f J_{x,y}$ in the case
of DA order $(f \ge 1/2)$.
To prove that the state in \eq{Factorized} is an eigenstate of the
Hamiltonian we decompose the latter
at $h=h_F$ into a sum of terms involving only pairs of
nn and nnn:
$H_{k,k+r} = f_r( J_x S_{k}^x S_{k+r}^x+ J_y S_{k}^y S_{k+r}^y
+J_z S_{k}^z S_{k+r}^z )  - h_{f}^r (S_{k}^z + S_{k+r}^z)$,
where, consistently with \eq{e.theta}, $h_{f}^r = f_r \cos\theta [J_z - \cos(\varphi_k)\cos(\varphi_{k+r})J_x]/2$.
Thus the condition for $\ket{G_F}$ to be an eigenstate
of every pair interaction term is:
\begin{equation}
\label{condition} - J_y + \cos^2\theta J_x  + \cos\varphi_k
\cos\varphi_{k+r} \sin^2 \theta J_z=0 \, .
\end{equation}
Because \eq{condition} must be satisfied both for the cases in which
$\varphi_k = \varphi_{k+r}$ and when $\varphi_k \neq \varphi_{k+r}$,
it must be either $\sin \theta=0$ or $J_z=0$. The first case ($\sin
\theta=0$) is trivial, as it implies saturation rather than proper
factorization. 
The second possibility
$(J_z=0)$ is associated with proper nontrivial factorization,
characterized by $\theta \neq 0$. Using \eq{condition} and
\eq{e.theta} one determines exactly the factorizing field:
\begin{equation}\label{field}
   h_F=\frac{1}{2}\sqrt{\mathcal{J}_x \mathcal{J}_y} = \left\{
   \begin{array}{ll}
    (1-f) \sqrt{J_x J_y} & f<1/2 \\
    f \sqrt{J_x J_y} & f \ge 1/2
    \end{array} \; .
    \right.
\end{equation}
A sufficient condition for $\ket{G_F}$ to be the GS is
that its projection over every pair of spins be the GS of the
corresponding pair Hamiltonian \cite{PrlLungo,KurmannRossignoli}.
This condition is never satisfied in the presence of frustration,
whose effects cannot be captured by quantities involving
only pairs of spins. The method must be generalized to include
minimal finite subsets of spins encompassing frustration. In the
case of $r_{\max}=2$, the minimal subset is any block of three
contiguous spins, tagged $k-1$, $k$, and $k+1$. The corresponding
triplet Hamiltonian term
$H_k=\frac{1}{2}H_{k-1,k}+\frac{1}{2}H_{k,k+1}+H_{k-1,k+1}$
includes all the different types of irreducible interactions
appearing in the model. Exactly at $h=h_F$ we have that $H=\sum_k
H_k$. Moreover, the projection of $\ket{G_F}$ over the Hilbert space
of the three spins $k-1$, $k$, and $k+1$ is an eigenstate of $H_k$.
Therefore, if one can show that the projection of $\ket{G_F}$ is the
GS of every three-body term $H_k$, factorization of the total GS is
proven. The analysis yields that: (i) if $J_y=0$, the factorized
state \eq{Factorized} is the GS of the systems at $h=h_F$ for {\em
all} values of the frustration $f \in [0,1]$; (ii) if $J_y\neq 0$,
the GS is factorized when $f$ lies below a critical value $f_c$:
\begin{equation}\label{fcritico}
f_c=\frac{1}{2} \frac{J_x-\sqrt{J_xJ_y}+J_y}{J_x+J_y} \; .
\end{equation}
To assess whether
for $f > f_c$ there may be still a region compatible with GS
factorization we consider a
partition of the Hamiltonian into blocks of more than three spins.
We define the sequence of operators
$(\tilde{H}^{(n)}_k=\sum_{\gamma=-n}^n H_{k+\gamma})$ which, for any
integer $n$, admit the $(2n+1)$-spin projection of  $\ket{G_F}$ as
their eigenstate, and whose lowest eigenvalue, in the limit of large
$n$, coincides with the GS energy of the total Hamiltonian $H$.
For every $n$, the eigenvalue of
$\tilde{H}^{(n)}$ associated to the factorized eigenstate is
$\varepsilon(n)=(2n-1){\cal{E}}_{F}$, where ${\cal{E}}_{F}$
is the energy density per site at $h=h_F$. Denoting by $\mu(n)$ the
minimum eigenvalue of $\tilde{H}^{(n)}$, we have that only if there
exists an integer $\bar{n}$ such that $\Delta(n) \equiv \mu(n) -
\varepsilon(n)$ vanishes for any $n > \bar{n}$, then the factorized
state is associated to the lowest eigenvalues of $\tilde{H}^{(n)}$,
and hence it is the GS of the total Hamiltonian $H$. By
studying $\Delta(n)$ as a function of $n$ one can determine exactly,
albeit numerically, the actual boundaries separating the occurrence
and the absence of GS factorization, as reported in \fig{facground}.
The exact threshold value $f_t$ lies just slightly above the
analytical lower bound $f_c$, \eq{fcritico}.
Summarizing, we have shown that for $f < f_t$, short-range
frustrated models admit an exact, separable GS
of the form \eq{Factorized} when the external magnetic field $h$
takes the value $h_F$ defined by \eq{field}. The associated magnetic order
is SA, while every factorized state associated to a DA order is always
an excited energy eigenstate; it becomes a GS only exactly at $J_y =
0$. Because the existence of factorized energy eigenstates is always
associated to a violation of the parity symmetry of the
magnetization along the direction of the external field
\cite{PrlLungo,KurmannRossignoli} the proof of the existence of factorized
GSs, even if it yields no direct information on the location of the
quantum critical points $h_c$, warrants the existence of quantum
phase transitions to an ordered phase, in frustrated models, as the
external field $h$ decreases and crosses $h_c$ (in the case that we
have illustrated, it is a transition to a SA order along $x$).
Moreover, since the factorizing field $h_F$ necessarily lies in the
ordered region, one can, at least, conclude that the factorizing
field anticipates the critical one from below: $h_F \le h_c$, a
behavior already evidenced in some frustration-free models
\cite{Verrucchi}. Exactly at $f=f_t$ the system undergoes a level
crossing, and hence a first order phase transition from the
twofold degenerate factorized GS \eq{Factorized} to a
twofold degenerate entangled GS state with complex long range
incompatible with factorization points because the latter are
necessarily associated to ferromagnetic or antiferromagnetic orders
along a fixed axis \cite{PrlLungo}.
This phenomenon identifies a frustration-driven
entangling-disentangling transition of the GS at
$h = h_F$ as $f$ crosses the critical threshold $f_t$
dividing the regimes of weak and strong frustration.

A phenomenological measure of frustration
in antiferromagnetic models is provided by $T_{CW}/T_N$ i.e. the ratio
of the Curie-Weiss temperature to the Neel temperature of bulk
three-dimensional ordering \cite{Balents}. This definition
cannot be applied to systems with vanishing $T_N$
(like, e.g. 1-D and 2-D models) and it cannot
distinguish between different contributions, classical and quantum,
to frustration. Notwithstanding these limitations,
our ground-state analysis suggests that this phenomenological measure
does capture some aspects of frustration, as follows. The existence of
a factorized GS implies the existence of a transition to an ordered phase
(occurring at a critical field higher than the  factorizing field $h_c > h_f$).
In a bulk 3-D system this should correspond to a finite value
of the ratio $T_{CW}/T_N$, since the order would necessarily freeze at a finite
value of $T_N > 0$. On the other hand, in strongly frustrated models, strong
correlations between quantum fluctuations
persist in the presence of an applied field favoring GS factorization,
and the existence of ordered phases tends to be suppressed.
In the corresponding bulk 3-D systems one should then find a higher,
asymptotically diverging value of the ratio $T_{CW}/T_N$, as the temperature at
which the order freezes approaches the absolute zero.

\smallskip

\begin{figure}[b]
\includegraphics[width=8cm]{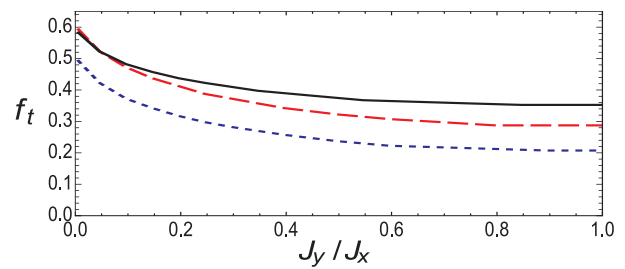}
\caption{(Color online) Threshold value of the frustration, $f_t$,
below which GS factorization occurs, for frustrated antiferromagnets with
$r_{\max}=4$, as a function of $J_y/J_x$. Solid black line:
$f_2=f$; $f_3=f^2$; $f_4=f^3$. Dashed red line: $f_2=f$; $f_3=f/2$;
$f_4=f/3$. Dotted blue line:  $f_2=f$; $f_3=f/2$; $f_4=f/4$.}
\label{facground-4}
\end{figure}

\noindent {\bf Models with interactions of arbitrary finite
range.--} For models \eq{HamRange} with finite
$r_{\max} > 2$ the triplet Hamiltonians $H_k$ are generalized
to subsets of $r_{\max}+1$ spins, with constraint $\sum_k H_k =H$
at $h = h_F$.
The space of the Hamiltonian parameters is still divided in a region
of low frustration compatible with GS factorization, and one of high
frustration for which GS factorization is forbidden, as shown in
\fig{facground-4} for models with maximum range of interaction
$r_{\max}=4$ that include, for instance, the $J_1 - J_2 - J_3 - J_4$
and $J_1 - J_2 - J_3$ models. Two general trends are
observed: For $J_y \neq 0$, the factorized GS has SA
order along the $x$-axis and, as shown in \fig{facground} and
\fig{facground-4}, the region of low frustration allowing GS
factorization decreases as the anisotropy $J_y/J_x$ increases.

\smallskip

\noindent {\bf Models with interactions of infinite range.--} If in
\eq{HamRange} we let $f_r \rightarrow 0$ when the maximum
interaction range $r_{max} \rightarrow \infty$, the question of the
existence of factorized energy eigenstates can be analyzed by
neglecting all the interactions between spins at distances greater
then some cutoff value $r'$, solve the associated constraints, and
then let $r' \rightarrow \infty$. To this aim, for each $r'$ we
consider the operator $H_k^{(r')}=\frac{1}{2} \sum_{\gamma=-r'}^{r'}
\sum_{\gamma ' =-r'}^{r'} (1-\delta_{\gamma-\gamma '})
\frac{1}{2r'+1-|\gamma-\gamma '|} H_{\gamma,\gamma '}$ that
expresses the sum of all the pair interaction terms between the $r'$
spins closest to $k$, and the associated quantity
$\Delta(r')=\mu(r')+\frac{1}{4}(J_x + J_y) \sum_{k=1}^{r'} (-1)^k
f_k$, where $\mu(r')$ is the lowest eigenvalue of $H_k^{(r')}$. We
have then analyzed different decay laws for $f_r$, i.e. fast:
$f_r=1/r^2$, intermediate: $f_r=1/r$, and slow: $f_r=1/\sqrt{r}$. In
the first case it is always $\Delta(r')=0$ and consequently the
system admits a factorized GS exactly at
$h_F=(\pi^2/12)\sqrt{J_xJ_y}$. In the second case ($f_r=1/r$),
according to the numerical evidence, $\Delta(r')$ vanishes in the
limit of arbitrarily large $r'$ and GS factorization appears to
occur at $h_F=\ln(2)\sqrt{J_xJ_y}$. Finally, in the case of slow
decay ($f_r=1/\sqrt{r}$), one has that $\Delta(r') \neq 0 \forall
r'$ and therefore no factorized GS can exist. Therefore, fully
connected models characterized by a rapidly decaying $f_r$, and hence
by low frustration, allow for GS factorization and the associated
SA or DA orders. Viceversa, models with slowly decaying $f_r$,
corresponding to strong frustration, do not admit factorized GS and
simple mean-field and classical-like descriptions.

\smallskip

\noindent {\bf Frustrated quantum models of complex condensed
matter and biological systems.--}
The ANNNI (Axial Next-Nearest-Neighbor Ising) model,
a particular case of the general class of models that we consider in the present work, provides
a possible effective description of systems with long-period modulated
structures, such as, e.g., polytypism, anti-phase boundaries in
binary alloys, and helical phases in rare earths compounds \cite{ANNNI}.
It is thought that quantum frustration effects may be
the mechanism responsible for the observed stability of these structures,
and for this reason the quantum version of the ANNNI model
is being intensively studied \cite{ANNNI}.
It is then important to establish whether stable modulated structures
are indeed predicted at all by the quantum ANNNI model and in what physical regimes.
When applied to the quantum ANNNI model ($J_y = J_z = 0$), our analysis
proves that the mean-field description is applicable for all values of
the frustration and that the value $f = 1/2$ discriminates between
two types of stable structures, a simple unmodulated ferromagnetic order
associated to a fully factorized GS for $f < 1/2$ and an anti-phase
modulated GS with DA order for $f > 1/2$.

Models of frustrated quantum spin networks have also been advocated as effective
descriptions of gene expression and complex genomic patterns \cite{Sasai}. Here,
again, the problem arises of the range of applicability of simple mean-field
descriptions corresponding to simple magnetic orders.
Indeed, much as for neural networks, the landscape of
stable attractors in gene networks depends, classically, on the degree of frustration.
Assuming a description based on frustrated classical models with long-range
interactions leads to the qualitative prediction of a small number of stable attractors
in the presence of a "sufficiently" weak frustration. The question is then whether
this prediction is stable against the effects of quantum fluctuations. Our analysis
shows that the mean-field picture is qualitatively correct and makes it quantitative by
determining the boundary between the weak and the strong frustration regime,
in which the mean-field predictions fail. For models
with long-range interactions, as we have shown above (see Fig. \fig{facground-4}),
the region of low frustration consistent with a mean-field description is determined
by the anisotropy ratio $J_y/J_x$ and decreases as the latter is increased.

\smallskip

\noindent {\bf Conclusions and outlook.--}
We have introduced a rigorous criterion for discriminating
between weakly and strongly frustrated quantum systems in terms of
GS factorizability. We have determined the threshold that
separates the regions of weak and strong frustration, and we have singled
out the exact forms of the factorized GS, the associated quantum phases,
and the corresponding magnetic orders in the region of low frustration.
These criteria should be experimentally testable in two- and three-body
correlation experiments.

\smallskip

\noindent {\bf Acknowledgments.--}
This work has been realized in the framework of the FP7 STREP Project HIP
(Hybrid Information Processing) of the European Union, Grant n. 221889.
We acknowledge financial support also from INFN through Iniziativa Specifica PG 62.

\end{document}